\documentstyle[epsfig,subeqn,wstwocl]{article}
\begin{document}
\bibliographystyle{unsrt}
\renewcommand{\thefootnote}{\fnsymbol{footnote}}
\arraycolsep1.3pt
%
%%%%%%%%%%% DEFINITION OF ABBREVIATIONS %%%%%%%%%%%%%%%%%%%%
%
\def \be{\begin{equation}}
\def \ee{\end{equation}}
\def \bea{\begin{eqnarray}}
\def \eea{\end{eqnarray}}
%
%%%%%%%%%%%%%%%%%%%%%%%%%%%%%%%%%%%%%%%%%%%%%%%%
%
\def \bem#1{\renewcommand{\thefootnote}{\arabic{footnote}}\footnote{#1}}
\def \bra#1{\left\langle #1\right|}
\def \braket#1#2#3{\left\langle #1\right|#2\left| #3\right\rangle}
\def \ket#1{\left| #1\right\rangle}
\def \cf{{\it cf.\/}}
\def \com#1#2{\left[#1,#2\right]}
\def \cp{{\it CP\/}}
\def \cpt{{\it CPT\/}}
\def \e{\mathrm{e}}
\def \ea{{\it et al.\/}}
\def\prs#1#2#3  {{\em Proc. Roy. Soc.} {\bf#1} (#2) #3.}
\def \eg{e.g.}
\def \etc{{\it etc.\/}}
\def \eq#1{Eq.~(\ref{#1})}
\def \eqs#1#2{Eqs.~(\ref{#1})--(\ref{#2})}
\def \fig#1{Fig.~\ref{#1}}
\def \figs#1#2{Figs.~\ref{#1}--\ref{#2}}
\def \fqm{{\mathrm FQM}}
\def \ha{Hamiltonian}
\def \half{{1\over 2}}
\def \heff{H_{\mathrm{eff}}}
\def \H{heavy quark effective theory}
\def \HQ{HQET}
\def \Im{{\mathrm{Im}}\,}
\def \ie{i.e.}
\def \la{Lagrangian}
\def \L{{\cal L}}
\def \leff{\L_{\mathrm{eff}}}
\def \lhqet{\L_{\mathrm{HQET}}}
\def \nnu{\nonumber}
\def \norm#1#2{\left\langle#1|#2\right\rangle}
\def \O{{\cal O}}
\def \ol#1{\overline{#1}}
\def \qb#1{\bar{#1}}
\def \ul#1{\underline{#1}}
\def \Re{{\mathrm{Re}}\,}
\def \rf{Ref.~\cite}
\def \rfs{Refs.~\cite}
\def \sec#1{Sec.~\ref{#1}}
\def \SI{\sum \limits_{X_s,\,{\mathrm{pol}}}}
\def \t+{\tau^+}
\def \t-{\tau^-}
\def \Tr{{\mathrm{Tr}}\,}
\def \Vec#1{\mbox{\bf#1}}
%
%%%%%%%%%%%%%%%%% GREEK LETTERS %%%%%%%%%%%%%%%%%%%%%%
%
\def \a{\alpha}
\def \b{\beta}
\def \g{\gamma}
\def \G{\Gamma}
\def \d{\delta}
\def \epsi{\epsilon}
\def \l{\lambda}
\def \La{\Lambda}
\def \m{\mu}
\def \n{\nu}
\def \p{\pi}
\def \r{\rho}
\def \s{\sigma}
\def \S{\Sigma}
\def \t{\tau}
%
%%%%%%%%%%%%%%%%%% DECAYS %%%%%%%%%%%%%%%%%%%%%%%%%
%
\def \bkll{$\B\to K l^+l-$}
\def \bk*ll{$\B\to K^* l^+l-$}
\def \bxll{$\B\to X_s l^+l-$}
\def \bktt{$\B\to K \tau^+\tau^-$}
\def \bk*tt{$\B\to K^* \tau^+\tau^-$}
\def \bxtt{$\B\to X_s \tau^+\tau^-$}
%
%%%%%%%%%%%%%%% DIMENSIONLESS QUANTITIES %%%%%%%%%%%%%%%%%
%
\def \msh{\hat{m}_s}
\def \mth{\hat{m}_{\t}}
\def \sh{\hat{s}}
%
%%%%%%%%%%%%% SLASH COMMANDS %%%%%%%%%%%%%%%%%%%%%%%%
%
\def \dslash{D\hspace{-0.65em} /}
\def \kslash{k\hspace{-0.50em} /}
\def \pslash{p\hspace{-0.42em} /}
\def \qslash{q\hspace{-0.45em} /}
\def \sslash{s\hspace{-0.45em} /}
\def \vslash{v\hspace{-0.45em} /}
%
%%%%%%%%%%%%%%%% JOURNALS %%%%%%%%%%%%%%%%%%%%%%%%%
%
\def\ap#1#2#3   {{\em Ann. Phys. (NY)} {\bf#1} (#2) #3.}
\def\apj#1#2#3  {{\em Astrophys. J.} {\bf#1} (#2) #3.}
\def\apjl#1#2#3 {{\em Astrophys. J. Lett.} {\bf#1} (#2) #3.}
\def\app#1#2#3  {{\em Acta. Phys. Pol.} {\bf#1} (#2) #3.}
\def\ar#1#2#3   {{\em Ann. Rev. Nucl. Part. Sci.} {\bf#1} (#2) #3.}
\def\cpc#1#2#3  {{\em Computer Phys. Comm.} {\bf#1} (#2) #3.}
\def\err#1#2#3  {{\it Erratum} {\bf#1} (#2) #3.}
\def\ib#1#2#3   {{\it ibid.} {\bf#1} (#2) #3.}
\def\jmp#1#2#3  {{\em J. Math. Phys.} {\bf#1} (#2) #3.}
\def\ijmp#1#2#3 {{\em Int. J. Mod. Phys.} {\bf#1} (#2) #3.}
\def\jetp#1#2#3 {{\em JETP Lett.} {\bf#1} (#2) #3.}
\def\jpg#1#2#3  {{\em J. Phys. G.} {\bf#1} (#2) #3.}
\def\mpl#1#2#3  {{\em Mod. Phys. Lett.} {\bf#1} (#2) #3.}
\def\nat#1#2#3  {{\em Nature (London)} {\bf#1} (#2) #3.}
\def\nc#1#2#3   {{\em Nuovo Cim.} {\bf#1} (#2) #3.}
\def\nim#1#2#3  {{\em Nucl. Instr. Meth.} {\bf#1} (#2) #3.}
\def\np#1#2#3   {{\em Nucl. Phys.} {\bf#1} (#2) #3.}
\def\pcps#1#2#3 {{\em Proc. Cam. Phil. Soc.} {\bf#1} (#2) #3.}
\def\pl#1#2#3   {{\em Phys. Lett.} {\bf#1} (#2) #3}
\def\prep#1#2#3 {{\em Phys. Rep.} {\bf#1} (#2) #3.}
\def\prev#1#2#3 {{\em Phys. Rev.} {\bf#1} (#2) #3}
\def\prl#1#2#3  {{\em Phys. Rev. Lett.} {\bf#1} (#2) #3.}
\def\prs#1#2#3  {{\em Proc. Roy. Soc.} {\bf#1} (#2) #3.}
\def\ptp#1#2#3  {{\em Prog. Th. Phys.} {\bf#1} (#2) #3.}
\def\ps#1#2#3   {{\em Physica Scripta} {\bf#1} (#2) #3.}
\def\rmp#1#2#3  {{\em Rev. Mod. Phys.} {\bf#1} (#2) #3.}
\def\rpp#1#2#3  {{\em Rep. Prog. Phys.} {\bf#1} (#2) #3.}
\def\sjnp#1#2#3 {{\em Sov. J. Nucl. Phys.} {\bf#1} (#2) #3.}
\def\spj#1#2#3  {{\em Sov. Phys. JEPT} {\bf#1} (#2) #3.}
\def\spu#1#2#3  {{\em Sov. Phys.-Usp.} {\bf#1} (#2) #3.}
\def\zp#1#2#3   {{\em Zeit. Phys.} {\bf#1} (#2) #3.}
%
%%%%%%%%% END OF DEFINITIONS %%%%%%%%%%%%%%%%%%%%%%%%%%%%
%
\setcounter{secnumdepth}{2}
%
%%%%%%%%%%%%%%%%%%%%%%%%%%%%%%%%%%%%%%%%%%%%%%%%
%
%%
%%                                                  %
%
%%     - PAPER -
%%                                    %
%
%%
%%                                                  %
%%%%%%%%%%%%%%%%%%%%%%%%%%%%%%%%%%%%%%%%%%%%%%%%
%
\title{REMARKS ON THE INCLUSIVE DECAYS
$\Lambda_b \to X_s\gamma$ AND $\Lambda_b\to X_c l\bar{\nu}_l$
\footnotemark}
\firstauthors{M. Gremm, G. K\"opp, F. Kr\"uger and L.\,M. Sehgal}
\firstaddress{Institut f\"ur Theoretische Physik (E), RWTH Aachen,
D-52074 Aachen, Germany}
\twocolumn[
\begin{flushright}
PITHA 95/28\par
\vskip 0.5em
hep-ph/9510273\par
\vskip 0.5em
October 1995\par
\vskip 2em
\end{flushright}
\faketitle\abstracts{
We consider the effects of quark-binding on the angular distribution and
polarization characteristics of the inclusive decays $\La_b\to X_s\g$ and
$\La_b\to X_c l \bar{\n}_l$ $(l=e, \t)$, using the methods of heavy quark
effective theory.}]

\footnotetext{Talk presented by L.\,M. Sehgal at the {\it{International
Europhysics Conference on
High Energy Physics}},
Brussels, 27 July--2 August, 1995. To appear in the proceedings.}

\section{Introduction}
This is a summary of two papers\cite{gks,ggs} in which we have considered
the effects of quark-binding on the decays $b\to s\g$ and $b\to q l
\bar{\n}_l$, when
the $b$-quark is embedded in a polarized $\Lambda_b$-baryon.

In the free-quark model (FQM), the decay $\stackrel{\Rightarrow}{b}\to s \g$ of
a polarized $b$-quark produces
a monochromatic photon whose angular distribution relative to the $b$-spin
direction is
\begin{subequations}\label{one}
\be\label{1a}
\frac{d\G}{d\cos{\theta}}\sim 1-\frac{1-\xi}{1+\xi}\cos{\theta}\ ,
\ee
where $\xi=m_s^2/m_b^2$. The polarization of the photon is
\be\label{1b}
P_{\g}=-\frac{1-\xi}{1+\xi}\ .
\ee
\end{subequations}
Likewise, in the decay $\stackrel{\Rightarrow}{b}\to q l \bar{\n}_l$ the
angular distribution of the lepton
relative to the $b$-spin is
\be\label{lepton1}
\frac{d\G}{d\cos{\theta}} \sim 1- \frac{1}{3}\,f\left(\frac{m_q^2}{m_b^2},
\frac{m_l^2}{m_b^2}\right)\cos{\theta}\ ,
\ee
where $f\left(m_q^2/m_b^2,m_l^2/m_b^2\right)$ is a calculable function
equal to unity when $m_q$ and $m_l$ are zero. The lepton in the final state has
a characteristic polarization $\vec{P}$, with a longitudinal component $P_L$
and a transverse component $P_T\sim m_l/m_b$ in the decay plane. Our objective
is to analyse how the free-quark characteristics (\ref{one}) and
(\ref{lepton1}) are
modified when the $b$-quark is a constituent of a polarized
$\Lambda_b$-baryon.

\section{The decay $\Lambda_b \to X_s\gamma$}
The decay $\stackrel{\Rightarrow}{\Lambda_b} \to X_s\gamma$ is governed by the
effective Hamiltonian
\bea\label{effham}
\heff&=&\frac{-4G_F}{\sqrt{2}}\frac{e}{16\pi^2}\,V_{tb}^{}V_{ts}^*
\,c_7(m_b)\nnu\\
&\times&\bar{s}\,\sigma^{\mu\nu}\left(m_b P_R+m_s P_L\right)b F_{\mu\nu}\ ,
\eea
which produces a distribution $(y=2E_{\g}/m_b)$
\be\label{rate1}
\frac{d\Gamma}{dy\,d\cos\theta}=\frac{\alpha G_F^2 m_b^2}{2^7 \pi^5}
\left|V_{tb}^{}V_{ts}^*\right|^2 \left|c_7(m_b)\right|^2 y\,\Im
T(y,\cos\theta)\ ,
\ee
where the function $T(y, \cos{\theta})$, calculated to order $1/m_b^2$ in the
operator product expansion (OPE), is\cite{gks}
\bea\label{amplitude1}
T(y,\cos\theta)&=&2y^2m_b^3\,\frac{1}{(y-y_0-i\epsilon)}\nnu\\
&\times&\left\{\left[1+ h(y) K\right](1+\xi)\right. \nnu \\
&-&\cos\theta\left.\left[1+\epsilon_b+h(y) K \right](1-\xi)\right\}\ ,
\eea
with
\be
h(y) = \frac{5}{3}
-\frac{7}{3}\,\frac{y}{(y-y_0-i\epsilon)}+\frac{2}{3}\,
\frac{y^2}{(y-y_0-i\epsilon)^2}\ .
\ee
The parameters $K$ and $\epsi_b$ are defined by
\begin{eqnarray}
K&=&-\braket{\Lambda_b}{\bar h_v\frac{(iD)^2}{2m_b^2}h_v}{\Lambda_b}\ ,\\
(1+\epsilon_b)s^{\mu}&=&\braket{\Lambda_b(s)}{\bar b\,\gamma^{\mu}\gamma^5b}
{\Lambda_b(s)}\ ,
\end{eqnarray}
and express the effects of quark-binding. The angular distribution of the decay
photon is
\begin{subequations}
\be\label{angular}
\frac{d\G}{d\cos{\theta}}\sim 1-K-(1+\epsi_b -K) \frac{1-\xi}{1+\xi}
\cos{\theta}\ ,
\ee
and the photon polarization, as a function of direction, is
\be\label{photonpola}
P_{\g}(\cos{\theta}) = -\frac{1-\xi-\alpha(1+\xi)\cos{\theta}}
{1+\xi-\alpha(1-\xi)\cos{\theta}}\ ,
\ee
\end{subequations}
with $\alpha = (1+\epsi_b-K)/(1-K)$. The results (\ref{angular}) and
(\ref{photonpola}), which are the QCD-improvements of the results
(\ref{1a}) and (\ref{1b}), are represented in Figs.~\ref{fractional} and
\ref{polarizationphoton}, for $K = 0.01$ and $\epsi_b = -\frac{2}{3} K$.
%
%%%%%%%%%% FIGURE 1 %%%%%%%%%%%%%%%%%%%%%%%%%%%%%%%%%
%
\setlength{\unitlength}{0.7mm}
\begin{figure}[p]
\begin{picture}(100,110)(0,1)
\mbox{\epsfxsize8.0cm\epsffile{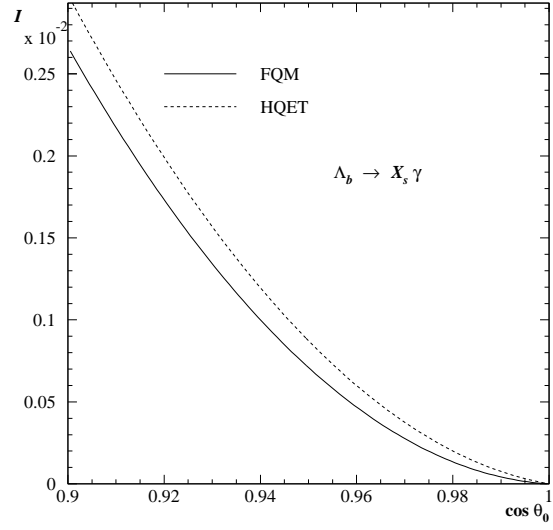}}
\end{picture}
\caption{The fractional intensity $I$ of photons in the forward cone
$\cos\theta_0\le\cos\theta\le 1$.\hfill}
\label{fractional}
\end{figure}
%
%%%%%%%%%% FIGURE 2 %%%%%%%%%%%%%%%%%%%%%%%%%%%%%%%%%
%
\setlength{\unitlength}{0.7mm}
\begin{figure}[p]
\begin{picture}(100,110)(0,1)
\mbox{\epsfxsize8.0cm\epsffile{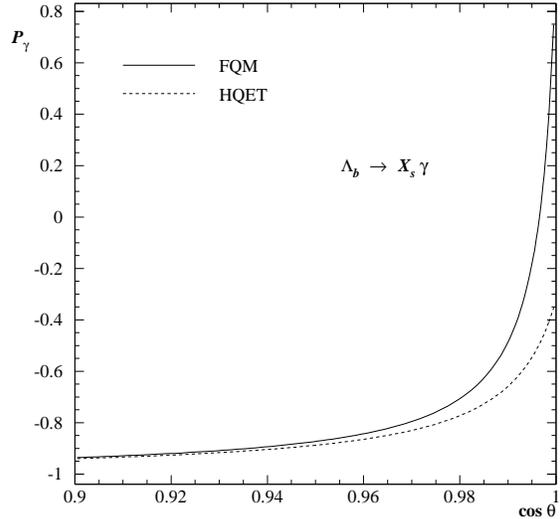}}
\end{picture}
\caption{The photon polarization $P_{\g}$ in the inclusive $\Lambda_b$
decay as a function of the photon direction with $K= 0.01$ and
$\epsilon_b=-\frac{2}{3}K$.}
\label{polarizationphoton}
\end{figure}
%
%%%%%%%%%%%%%%%%%%%%%%%%%%%%%%%%%%%%%%%
%
\section{The decay $\La_b\to X_q l \bar{\n}_l$}
In the case of the decay $\stackrel{\Rightarrow}{\La_b}\to X_q l \bar{\n}_l$
$(q=u$ or $c$, $l= e$ or $\t)$,
the differential decay rate has the form $d\G \sim L_{\m\n}H^{\m\n}$, where
the hadronic tensor is
\be\label{hadrontensor}
H_{\mu\nu}= \frac{1}{\p}\Im i \int d^4x\,\,e^{-iq\cdot x}\braket{\Lambda_b}
{{\mathrm T}\left\{j_{\m}^{\dagger}(x) j_{\n}^{}(0)\right\}}{\Lambda_b}\ ,
\ee
with $j_{\m} = V_{qb}\,\bar{q}\g_{\m}(1-\g_{5})b$. This tensor may be expanded
in terms of 5 structure functions $T_1\ldots T_5$, for unpolarized $\La_b$,
and 9 additional functions $S_1\ldots S_9$, for polarized $\La_b$. Using OPE
to order $1/m_b^2$, it is possible to determine all of these structure
functions
in terms of the parameters $K$ and $\epsi_b$.

The inclusive distribution of the lepton as a function of energy and angle has
the
form\cite{ggs} $(y = 2 E_l/m_b)$
\be\label{leptondistr}
\frac{d\Gamma}{dy\,d\cos\theta} \sim \left[A_0 + A_1 K
+ \left\{B_0(1+\epsi_b)+B_1 K \right\} \cos{\theta} \right]\ ,
\ee
where $A_{0,1}$ and $B_{0,1}$ are calculable functions of $y$. In the limiting
case $\La_b\to X_u e \bar{\n}_e$, with $m_u = m_e = 0$, the functions $A_0$,
$B_0$ have a form familiar from $\m$-decay:
\be\label{A0B0}
A_0 = (3-2 y) y^2\ , \quad\quad B_0 = (1-2y) y^2\ .
\ee
Table \ref{table1} lists the inclusive distribution $d\G/d\cos{\theta}$,
integrated over $y$, for the various inclusive processes
$\stackrel{\Rightarrow}{\La_b}\to X_q l \bar{\n}_l$.
%
%%%%%%%%%% TABLE 1 %%%%%%%%%%%%%%%%%%%%%%%%%%%%%%%%%
%
\begin{table}
%\twocolumn[\caption{Forward-Backward Asymmetry (integrated over $y$)}]
\begin{center}\caption{Angular distribution of leptons in the
inclusive decay $\stackrel{\Rightarrow}{\La_b}\to X_q l \bar{\n}_l$, including
quark-binding effects.}\label{table1}
\vspace{0.3cm}
\begin{tabular}{c | c }
\hline\hline
&\\
Decay & $d\G /{\cos{\theta}}$ (arbitrary unit)\\ \hline
&   \\
$\La_b\to X_u e \bar{\n}_e$ & $(1-K)-\frac{1} {3}(1+\epsi_b-K)\cos{\theta}$ \\
& \\
$\La_b\to X_c e \bar{\n}_e$ &  $(1-K)-0.25(1+\epsi_b-K)\cos{\theta}$   \\
& \\
$\La_b\to X_u \t \bar{\n}_{\t}$ &  $(1-K)-0.45(1+\epsi_b-1.8\,K)\cos{\theta}$
\\
& \\
$\La_b\to X_c \t \bar{\n}_{\t}$ &  $(1-K)-0.34(1+\epsi_b-2.7\,K)\cos{\theta}$
\\ \hline\hline
\end{tabular}
\end{center}
\end{table}
%
%%%%%%%%%%%%%%%%%%%%%%%%%%%%%%%%%%%%%
%
A further observable of interest is the $\t$-polarization in the decay
$\La_b\to X_c \t \bar{\n}_{\t}$. The $\t$ has a longitudinal polarization
component $P_L$ as well as a transverse component $P_T$ in the decay plane.
These are shown in Figs.~\ref{longpol} and \ref{transpol}, as functions of
the lepton energy. The average values are $\left\langle P_L\right\rangle
\approx -0.70$ and $\left\langle P_T\right\rangle  \approx 0.19$.
The quark-binding effects are generally small, except
in the region of large $y$, where the OPE breaks down, and a ``smoothing''
procedure is necessary. When the $\La_b$ is polarized, there is an interesting
correlation of $\vec{s}_{\t}$ with $\vec{s}_{\La_b}$. In particular, the
$\t$-lepton can have a small polarization component $P_{\perp}$ perpendicular
to the decay plane, which is proportional to $K$, and hence a pure
manifestation
of the ``Fermi motion'' of the $b$-quark in the hadron.
%
%%%%%%%%%% FIGURE 3 %%%%%%%%%%%%%%%%%%%%%%%%%%%%%%%%%
%
\setlength{\unitlength}{0.7mm}
\begin{figure}[p]
\begin{picture}(100,110)(0,1)
\mbox{\epsfxsize8.0cm\epsffile{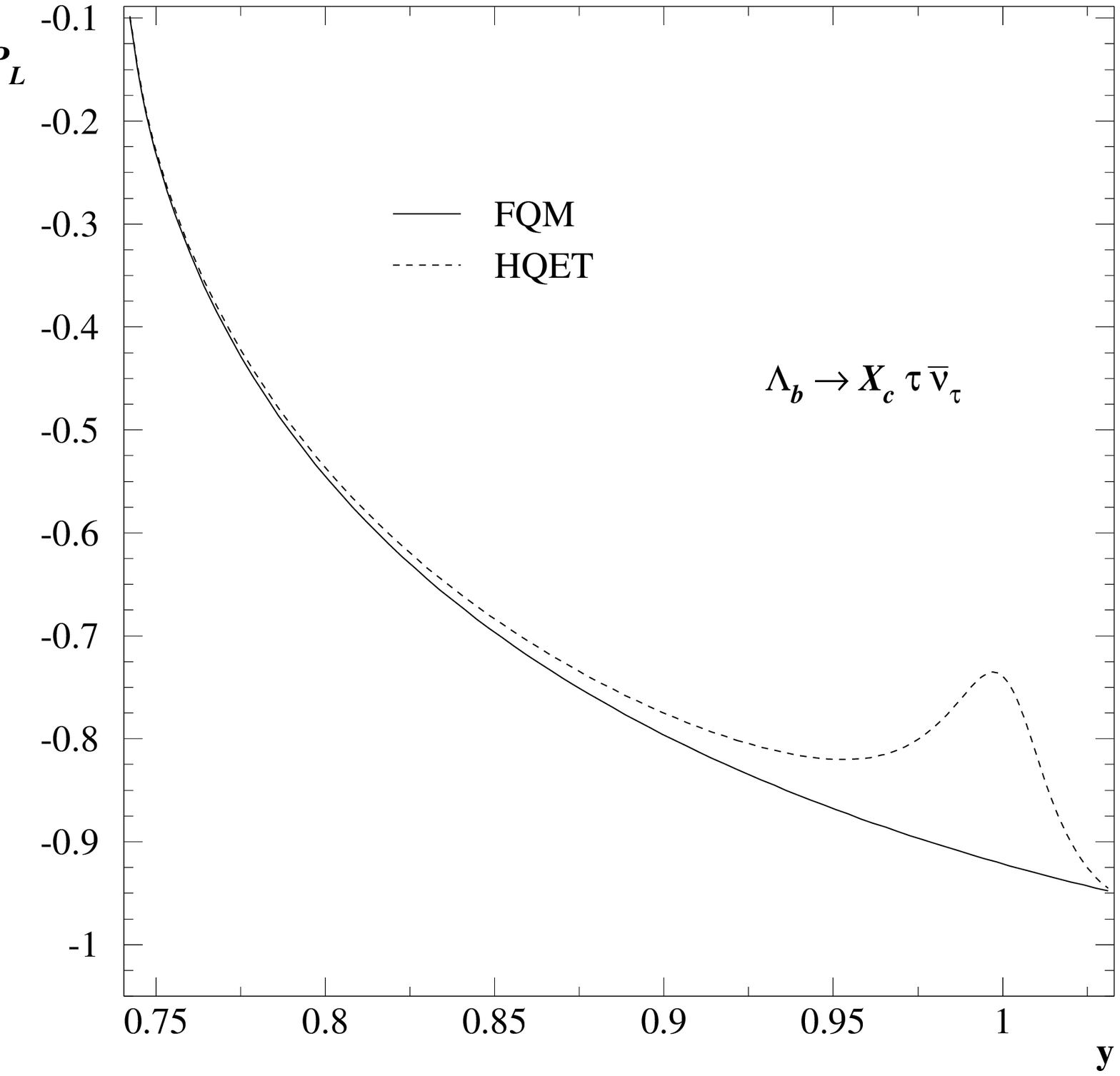}}
\end{picture}
\caption{Longitudinal polarization of $\t$ in $\La_b\to X_c\t\bar{\n}_{\t}$.}
\label{longpol}
\end{figure}
%
%%%%%%%%%% FIGURE 4 %%%%%%%%%%%%%%%%%%%%%%%%%%%%%%%%%
%
\setlength{\unitlength}{0.7mm}
\begin{figure}[p]
\begin{picture}(100,110)(0,1)
\mbox{\epsfxsize8.0cm\epsffile{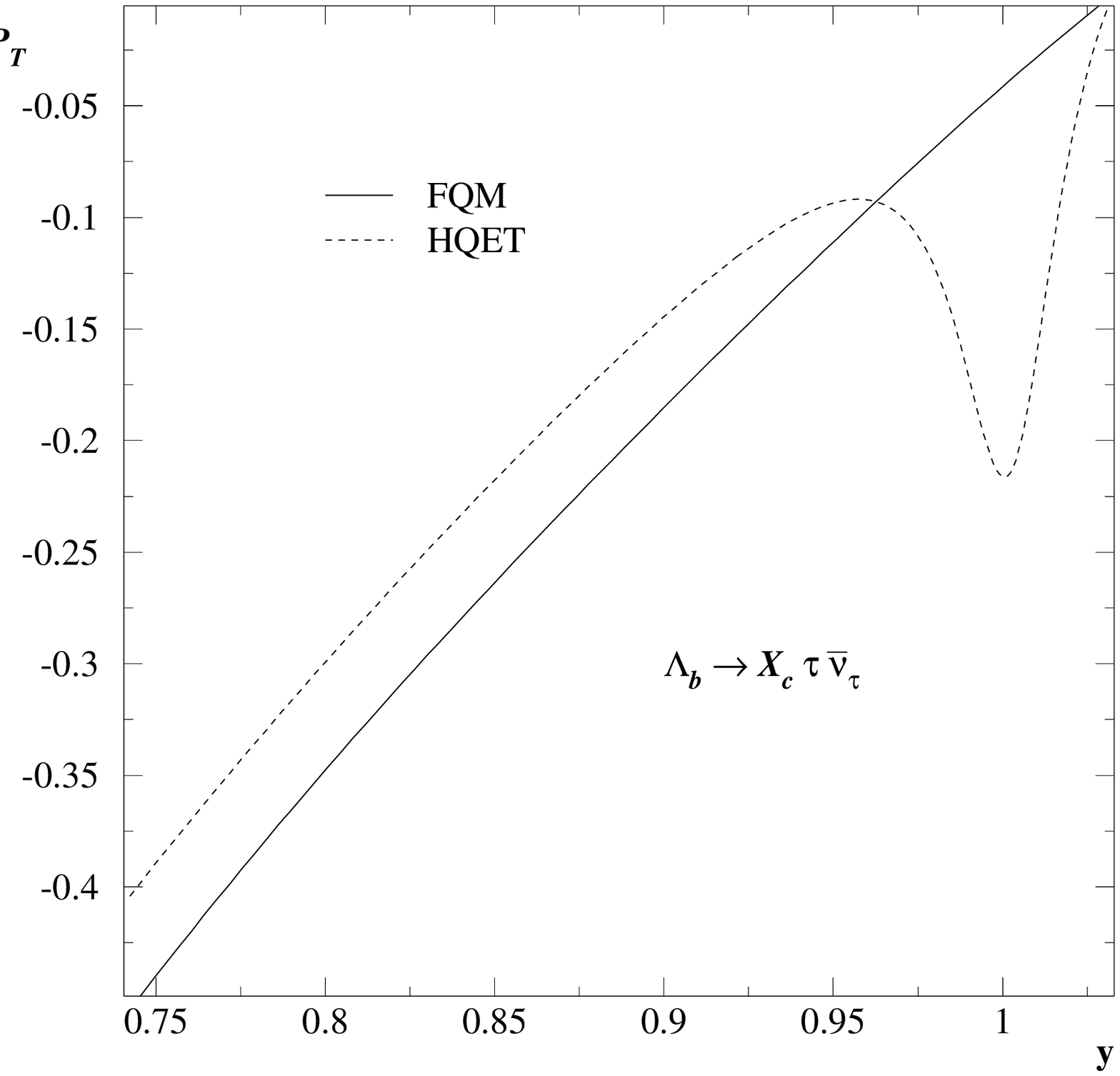}}
\end{picture}
\caption{Transverse polarization of $\t$ (in decay plane) in $\La_b\to
X_c\t\bar{\n}_{\t}$.}
\label{transpol}
\end{figure}
%
%%%%%%%%%%%%%%%%%%%%%%%%%%%%%%%%%%%%%%%%%%%%%%
%
\setcounter{secnumdepth}{0}

%\newpage

\end{document}